\newif\ifarXiv
\arXivtrue   
\ifarXiv
\documentclass[12pt]{article}
\usepackage{epsfig,amsmath,amsfonts,amssymb,makeidx,ifthen}
\usepackage[dvips]{color}
\usepackage{psfrag}
\usepackage[titletoc, title]{appendix}
\newtheorem{theorem}{Theorem}[section]
\newtheorem{lemma}[theorem]{Lemma}

\makeatletter
\@addtoreset{equation}{section}
\makeatother
\renewcommand{\theequation}{\thesection.\arabic{equation}}
\makeatletter
\long\def\@makecaption#1#2{{\small
\advance\leftskip1cm
\advance\rightskip1cm
\vskip\abovecaptionskip
\sbox\@tempboxa{#1: #2}%
\ifdim \wd\@tempboxa >\hsize
 #1: #2\par
\else
\global \@minipagefalse
\hb@xt@\hsize{\hfil\box\@tempboxa\hfil}%
\fi
\vskip\belowcaptionskip}}
\makeatother
\else
\documentclass[smallextended]{svjour3}  
\usepackage{graphicx,amsmath,amsfonts,amssymb,makeidx,ifthen,color}
\usepackage{psfrag,appendix}
\fi
\ifarXiv
\newcommand{\jour}[2]{#1}
\else
\newcommand{\jour}[2]{#2}
\fi
\newcommand{\calC}{\mathcal{C}}
\newcommand{\calD}{\mathcal{D}}
\newcommand{\calE}{\mathcal{E}}

\newcommand{\calH}{\mathcal{H}}
\newcommand{\calI}{\mathcal{I}}
\newcommand{\calJ}{\mathcal{J}}
\newcommand{\calK}{\mathcal{K}}

\newcommand{\calP}{\mathcal{P}}

\newcommand{\calV}{\mathcal{V}}

\newcommand{\up}{\uparrow}
\newcommand{\dn}{\downarrow}

\newcommand{\cxs}{c_{x,\sigma}}
\newcommand{\cxsd}{c_{x,\sigma}^\dagger}
\newcommand{\nxs}{n_{x,\sigma}}
\newcommand{\nxup}{n_{x,\uparrow}}
\newcommand{\nxdn}{n_{x,\downarrow}}
\newcommand{\aas}{a_{\alpha,\sigma}}
\newcommand{\aasd}{a_{\alpha,\sigma}^\dagger}
\newcommand{\bps}{b_{p,\sigma}}
\newcommand{\bpsd}{b_{p,\sigma}^\dagger}
\newcommand{\dps}{d_{p,\sigma}}
\newcommand{\dpsd}{d_{p,\sigma}^\dagger}
\newcommand{\acom}[1]{\{#1\}}
\newcommand{\Hhop}{H_\mathrm{hop}}
\newcommand{\Hint}{H_\mathrm{int}}
\newcommand{\halph}{h_\alpha}
\newcommand{\La}{\Lambda}
\newcommand{\Laa}{\Lambda_\alpha}

\newcommand{\bond}[1]{e(#1)}
\newcommand{\dLao}{\partial\Lambda_0}
\newcommand{\sumsigma}{\sum_{\sigma=\uparrow,\downarrow}}
\newcommand{\sumIud}{\sum_{I_\up,I_\dn\subset {\calI}}}
\newcommand{\sumIudo}{\sum_{I_\up,I_\dn\subset {\calI}\backslash\{0\}}}
\newcommand{\sumJo}{\sum_{k\in\calJ\backslash\{0\}}}
\newcommand{\sumtwo}[2]%
{\mathop{\sum_{#1}}_{#2}}
\newcommand{\rme}{\mathrm{e}}
\newcommand{\rmi}{\mathrm{i}}
\newcommand{\Ne}{N_\mathrm{e}}
\newcommand{\Stot}{S_\mathrm{tot}}
\newcommand{\vecS}{\mbox{\boldmath $S$}}
\newcommand{\PhiG}{\Phi_\mathrm{G}}
\newcommand{\tg}[2]{{\tilde{g}(#1,I_\up;#2,I_\dn)}}
\newcommand{\g}[2]{{{g}(\{#1\}\cup I_\uparrow;\{#2\}\cup I_\downarrow)}}
\begin{document}
\jour{ 
\noindent   
\textbf{\Large%
Ferromagnetism in the Hubbard model with 
a gapless nearly-flat band
}\bigskip\\
Akinori Tanaka%
\footnote
{%
Department of General Education, National Institute of Technology, 
Ariake College, Omuta, Fukuoka 836-8585, Japan\\
E-mail: akinori@ariake-nct.ac.jp
}
} 
{ 

\title{Ferromagnetism in the Hubbard model with a gapless nearly-flat band}

\titlerunning{Ferromagnetism in the Hubbard model with a gapless nearly-flat band}

\author{Akinori Tanaka}
%


\institute{A. Tanaka \at
              Department of General Education,
	      National Institute of Technology, Ariake College, Omuta, 
	      Fukuoka 836-8585, Japan \\
              Tel.: +81-944-53-8663\\
              Fax: +81-944-53-8663\\
              \email{akinori@ariake-nct.ac.jp}           
}

\date{Received: date / Accepted: date}

\maketitle
} 
%
%

\begin{abstract}
We present a version of the Hubbard model 
with a gapless nearly-flat lowest band
which exhibits ferromagnetism in two or more dimensions.
The model is defined on a lattice obtained by placing a site
on each edge of the hypercubic lattice,
and electron hopping is assumed to be only between 
nearest and next nearest neighbor sites. 
The lattice, where all the sites are identical, is simple,
and 
the corresponding single-electron band structure,
where 
two cosine-type bands touch  without an energy gap,
is also simple. 
We prove that the ground state of the model is unique and ferromagnetic
at half-filling of the lower band, 
if the lower band is nearly flat and the strength of 
on-site repulsion is larger than a certain value which
is independent of the lattice size. 
This is the first example of ferromagnetism in three dimensional non-singular
models with a gapless band structure. 

\jour{}
{
}
\end{abstract}
%
\jour{
\tableofcontents
}
{}
\section{Introduction}
Since the Hubbard model, a simple tight-binding model with
on-site repulsion,
was formulated in the early 1960s,
numerous attempts have been made on it to understand  
mechanisms for itinerant-electron ferromagnetism \cite{Lieb95,Tasaki98a,Tasaki98b}.
To date, some rigorous examples of ferrimagnetism \cite{Lieb89}, 
ferromagnetism \cite{Nagaoka66,Mielke91a,Mielke91b,Mielke92,Tasaki92,Mielke93,MielkeTasaki93,Tasaki95,Tasaki2003,TanakaUeda2003,Lu2009} 
and metallic ferromagnetism \cite{TanakaTasaki2007,TanakaTasaki2016} 
in the Hubbard model have been proposed, 
and it is well recognized that an interplay between
the quantum mechanical motion of electrons and repulsive 
interaction between electrons does generate ferromagnetism.

Among the others, flat-band models 
proposed by Mielke and Tasaki
provided a significant breakthrough in understanding of itinerant-electron
ferromagnetism \cite{Mielke91a,Mielke91b,Mielke92,Tasaki92,Mielke93,MielkeTasaki93}.
The flat-band models in common have multi single-electron bands
including a flat (highly degenerate) lowest band.
In these models, one finds that 
electrons occupy the flat band in order to minimize 
the kinetic energy 
and then
the spins of electrons occupying the flat band
align parallel to each other, i.e., align ferromagnetically, 
in order to avoid an increase of energy
due to on-site repulsion. 
The next important step was also taken by Tasaki, who proved
that the ferromagnetism in flat-band models on lattices constructed
by the cell-construction method 
is stable against perturbations which turn flat bands into
dispersive ones
provided that the on-site repulsion is sufficiently large \cite{Tasaki95,Tasaki2003}.
Note that
an electron has a tendency to occupy
a lower kinetic energy state when the repulsion is small.
Tasaki's models which have dispersive bands thus exhibit
the Pauli paramagnetism when the on-site repulsion is vanishing, 
and remain nonferromagnetic when the on-site repulsion is small.
Tasaki's models could describe the true competition between
the kinetic energy and the on-site repulsion, 
and established that the spin-independent repulsion 
can cause the itinerant-electron ferromagnetism.   

Although the stability of ferromagnetism in Tasaki's models 
is shown in rather general settings, 
as for Mielke's flat-band models on line graphs, 
less is known about stability or instability of ferromagnetism
in perturbed nearly-flat-band models.
Here we note that
there are no band gaps above flat lowest bands
in Mielke's flat-band models,
whereas
there are finite band gaps in
Tasaki's models and corresponding (unperturbed) flat-band models. 
This is an essential difference.
The occurrence of ferromagnetism in Tasaki's models,
at least at a heuristic level, 
can be understood as a consequence 
of the band gap as follows.
The band gap enforces the electrons to occupy 
the lowest nearly-flat band while the on-site repulsion 
forbids double occupancy
of sites. 
Then, despite that the lowest band is dispersive,
the situation is almost as in the flat-band models,
and the system exhibits ferromagnetism.
In fact, 
a certain parameter in hopping amplitudes which controls the energy gap 
in the band structure plays
an important role 
in the rigorous proof of ferromagnetism in Tasaki's models.
On the other hand, 
it is more subtle and difficult 
to show the stability of ferromagnetism against a perturbation
in Mielke's flat-band models
since there might be various low energy excitation modes 
reflecting a gapless nature of the band structure.
Some examples of nearly-flat-band models related to Mielke's flat-band models
in two dimensions, such as models on the kagome lattice and 
on the regular lattice of corner sharing tetrahedra \cite{TanakaUeda2003,Lu2009},
have been proposed and proved to have ferromagnetic ground states;
as far as we know, however, there are no examples 
in three or more dimensions.
  
In this paper, we propose a version of the Hubbard model which 
has a gapless nearly-flat lowest band
and exhibits ferromagnetism in the ground state in two or more dimensions. 
Our model is defined on a lattice obtained by placing
a site on each edge of the hypercubic lattice. 
In a certain case, it is reduced 
to a gapless flat-band model. 
In two dimensions the corresponding flat-band model
is Mielke's flat-band model on the regular lattice of
corner sharing tetrahedra.
In three or more dimensions, the corresponding flat-band
model is slightly different from Mielke's one, 
but they are closely related.
Although the perturbative hopping term considered in this paper 
is rather special,
our model is simple such that all the lattice sites are identical
and electron hopping is assumed to be only between nearest 
and next nearest neighbor sites.
We believe that our model is helpful for a better understanding
of ferromagnetism in the Hubbard model. 
\section{Definition of the model and the main result}
\label{s:Def}
\subsection{Definition of the lattice}
\label{subs:Def lat} 
We start by describing the lattice~$\La$ on which our
Hubbard Hamiltonian is defined.
Let $G=(\calV,\calE)$ 
be a graph, 
where $\calV$ is a set of vertices and $\calE$ is a set of edges.
The vertex set $\calV$ is assumed to be  
a subset of $\Bbb{Z}^\nu$ with $\nu\ge2$,
\begin{equation}
 \calV=\left\{ \alpha=(\alpha_1,\dots,\alpha_\nu)~|~
        \alpha_l\in\Bbb{Z}, 
       -\frac{L-1}{2} \le \alpha_l \le \frac{L-1}{2} 
        \mbox{ for $l=1,\dots,\nu$}
   \right\} \, ,
\end{equation}
where  $L$ is an odd integer%
\footnote{%
The reason why we choose $L$ to be an odd integer is
only technical. 
We can treat the case of even $L$ with a slight modification.}.
We impose periodic boundary conditions in all directions%
\footnote{%
This condition is also imposed to simplify the argument.
It is easy to apply the present method to the model 
with open boundary conditions.}.
Let $e(\alpha,\beta)$ be a line segment between 
nearest neighbor vertices~$\alpha$ and $\beta$ in $\calV$.
The edge set $\calE$ is a collection of these line segments%
\footnote{%
For $\alpha$ in $\calV$ or $x$ in $\La$ defined by~\eqref{eq:Lambda}, 
$|\alpha|$ or $|x|$ denotes the usual Euclidean distance.
For a set $X$, 
we use the same symbol $|X|$ to denote the number of
elements in $X$.},
\begin{equation}
 \calE=\left\{ \bond{\alpha,\beta}~|~\alpha,\beta\in {\calV}, 
                                    |\alpha-\beta|=1
   \right\}\, .
\end{equation}
For each edge~$\bond{\alpha,\beta}$ in $\calE$, 
we denote by $m(\alpha,\beta)$ the point taken  
in the middle of $\alpha$ and $\beta$. 
Then we define the lattice~$\La$ as a collection of these mid-points 
\begin{equation}
\label{eq:Lambda}
 \La = \{ x = m(\alpha,\beta)~|~\bond{\alpha,\beta}\in {\calE}\}\, .
\end{equation}

\begin{figure}
\begin{center}
\includegraphics[scale=1]{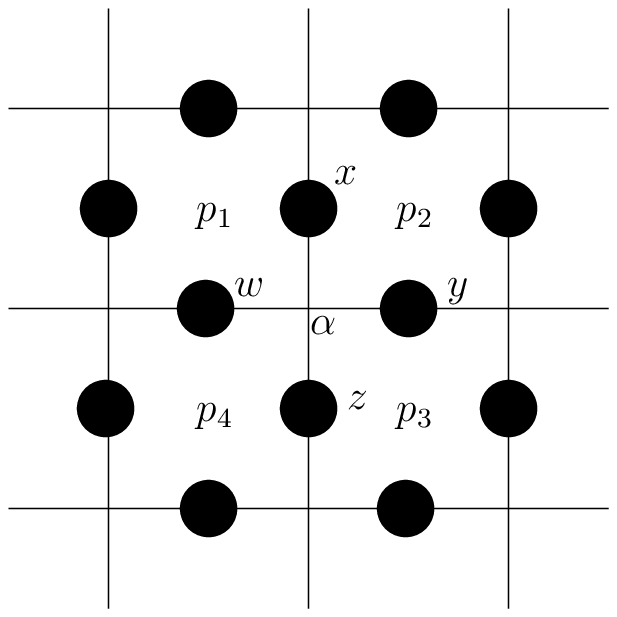}
\end{center}
\caption{The lattice structure for $\nu=2$. 
Thin lines represent edges in $\calE$ 
and filled circles represent lattice sites.
For example, the subsets 
$\calP(x)$ and  
$\calP_\alpha$ of $\calP$
are given by
$\calP(x)=\{p_1,p_2\}$ and 
$\calP_\alpha = \{p_1,p_2,p_3,p_4\}$, respectively.
The subset $\calC_\alpha$ of the lattice sites is
$\{w,x,y,z\}$.}
\label{fig:set of squares}
\end{figure}

Before proceeding to the definition of our Hubbard Hamiltonian,
let us introduce some more notation.
Let $\calP$ be the set of all regular squares with side-length 1
whose corners are located at vertices in $\calV$.
For four edges of a regular square $p$ in $\calP$, 
we find the corresponding mid-points,
sites in $\La$, which we denote by $x_i(p)$ with $i=1,2,3,4$.
For convenience, we assume that $x_3(p)$ denotes
the site opposite to that denoted by $x_1(p)$.  
Each edge whose mid-point is $x$ is shared by $2(\nu-1)$ 
regular squares in $\calP$;
we denote by $\calP(x)$ the collection of these regular squares. 
Each vertex $\alpha$ in $\calV$ is shared by $2\nu(\nu-1)$ regular
squares in $\calP$;
we denote by $\calP_\alpha$ the collection of
these regular squares.
Each vertex $\alpha$ in $\calV$ is shared by $2\nu$ edges,
$\bond{\alpha,\alpha+\delta_l}$ and
$\bond{\alpha,\alpha-\delta_l}$ with $l=1,\dots,\nu$,
where $\delta_l$ denotes the unit vector along the $l$-axis.
We define $\calC_\alpha$ as the collection of the mid-points of
these edges
\begin{equation}
 \calC_\alpha = \{ x = m(\alpha,y)~|~y=\alpha\pm\delta_l, l=1,\dots,\nu\}\, .
\end{equation}
The lattice structure for $\nu=2$ is shown in Fig.~\ref{fig:set of squares}. 
\subsection{Definition of the Hamiltonian
and the main result}
\label{subs:Def Ham}
Let $\cxs$ and $\cxsd$ be annihilation and creation 
operators, respectively, of an electron with spin~$\sigma$ 
at a site~$x$ in $\La$.
These operators satisfy the anticommutation relations 
\begin{equation}
\acom{c_{x,\sigma},c_{y,\tau}} = \acom{c_{x,\sigma}^\dagger,c_{y,\tau}^\dagger}=0
\end{equation}
and
\begin{equation}
\acom{c_{x,\sigma}^\dagger,c_{y,\tau}} = \delta_{x,y}\delta_{\sigma,\tau}
\end{equation}
for $x,y\in\La$ and $\sigma,\tau=\up,\dn$.
The number operator is defined by $\nxs = \cxsd\cxs$.

The total spin operators 
$\vecS_\mathrm{tot}=(\Stot^{(1)},\Stot^{(2)},\Stot^{(3)})$ 
of the system are defined by
\begin{equation}
 \Stot^{(i)} = \frac{1}{2}
              \sum_{x\in\La}
              \sum_{\sigma,\tau = \up,\dn}
              c_{x,\sigma}^\dagger
	      \mathsf{p}_{\sigma\tau}^{(i)}
	      c_{x,\tau}
\end{equation} 
for $i=1,2,3$, where $\mathsf{p}^{(i)}=[\mathsf{p}_{\sigma\tau}^{(i)}]_{\sigma,\tau=\up,\dn}$
are the Pauli matrices 
\begin{equation}
 \mathsf{p}^{(1)}=\left(\begin{array}{cc}
	        0 & 1 \\
                1 & 0 
                \end{array}
         \right)\,,\hspace*{1em}
 \mathsf{p}^{(2)}=\left(\begin{array}{cc}
	        0 & -\rmi \\
                \rmi & 0 
                \end{array}
         \right)\,,\hspace*{1em}
 \mathsf{p}^{(3)}=\left(\begin{array}{cc}
	        1 & 0 \\
                0 & -1 
                \end{array}
         \right)\,.
\end{equation}
We also define the raising and lowering operators on
the eigenvalues of $\Stot^{(3)}$ by
${\Stot^{+}=\Stot^{(1)}+\rmi\Stot^{(2)}}$ 
and ${\Stot^{-}=\Stot^{(1)}-\rmi\Stot^{(2)}}$,
respectively.  
We denote by $\Stot(\Stot+1)$ the eigenvalue of $(\vecS_\mathrm{tot})^2$.

We consider the Hubbard Hamiltonian on the lattice $\La$ given by
\begin{equation}
 \label{eq:Hamiltonian}
  H = \Hhop + \Hint\, ,
\end{equation}
where the first term in the right-hand side,
\begin{equation}
 \Hhop = \sumsigma
      \sum_{x,y\in\La} t_{x,y}c_{x,\sigma}^\dagger c_{y,\sigma}  
\end{equation}
with real parameters $t_{x,y}$, which we will specify below,
represents energy associated with 
the quantum mechanical motion of electrons,
and the second term,
\begin{equation}
 \Hint = U \sum_{x\in\La} \nxup\nxdn  
\end{equation}
with $U>0$, represents a repulsive interaction between
electrons with up- and down-spin at
the same site. 

\begin{figure}
(a)
\begin{center}
\includegraphics[scale=1]{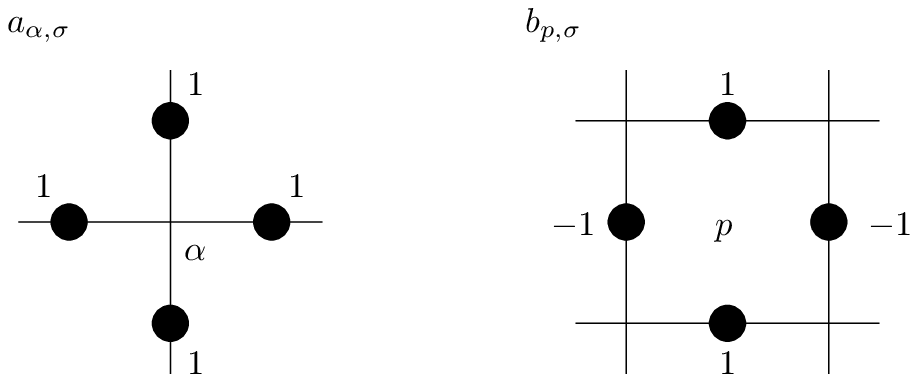}
\end{center}
(b)
\begin{center}
\includegraphics[scale=1]{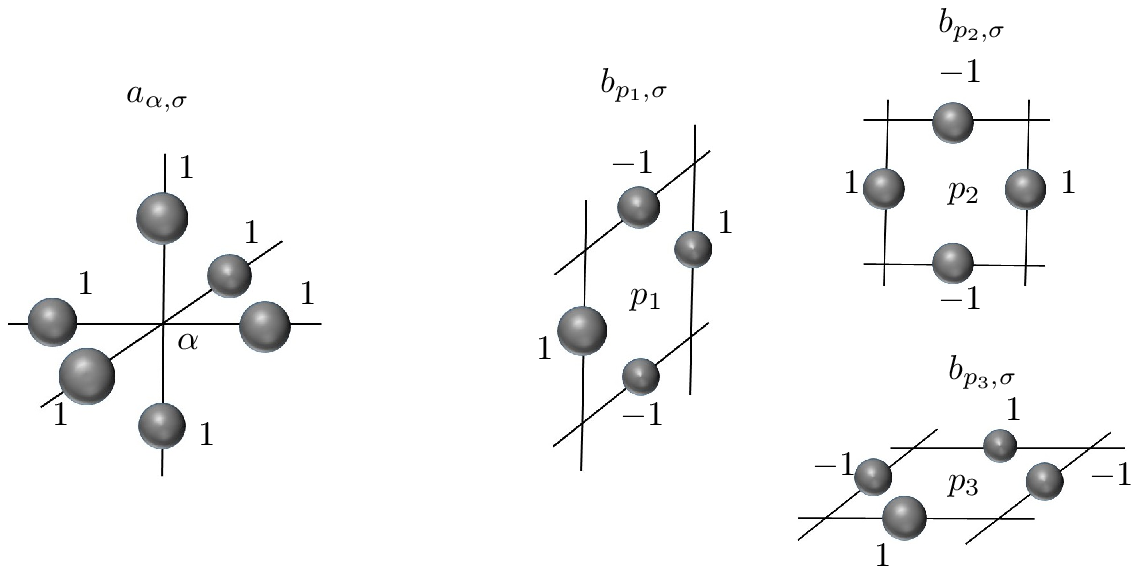}
\end{center}
\caption{The states corresponding to $a$- and $b$-operators.
Thin lines represent edges in $\calE$.
(a) $\nu=2$. (b) $\nu=3$. }
\label{fig:ab states}
\end{figure}

\begin{figure}[t]
\begin{center}
\includegraphics[scale=1]{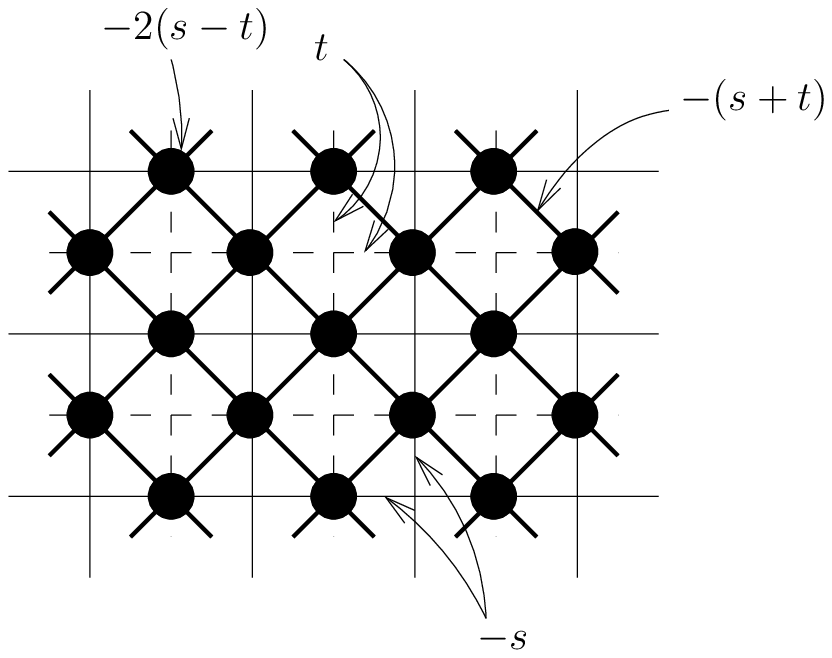}
\end{center}
\caption{The hopping amplitudes for $\nu=2$.
Filled circles represent lattice sites.}
\label{fig:lattice2d}
\end{figure}

The parameter $t_{x,y}$, which is called the hopping amplitude, is defined
as follows.
First we introduce new fermion operators. 
For each vertex $\alpha$ in $\calV$, let
\begin{equation}
 \aas = \sum_{x\in \calC_\alpha} c_{x,\sigma}\,,
\end{equation}
and for each regular square $p$ in $\calP$, let
\begin{equation}
 \bps = \sum_{i=1}^4(-1)^i c_{x_i(p),\sigma}
\end{equation} 
(see Fig.\ref{fig:ab states}).
We note that the following anticommutation relations are satisfied:
\begin{equation}
\label{eq:anticom aa}
 \acom{a_{\alpha,\sigma}^\dagger,a_{\beta,\sigma}}
  =\left\{ \begin{array}{ll}
            2\nu & \mbox{if $\alpha=\beta$};\\
            1 & \mbox{if $|\alpha-\beta|=1$};\\
            0 & \mbox{otherwise}\, ,
	   \end{array}
  \right.
\end{equation}
and 
\begin{equation}
 \acom{b_{p,\sigma}^\dagger,b_{q,\sigma}}
  =\left\{ \begin{array}{ll}
            4 & \mbox{if $p=q$};\\
            \mu[p,q] & \mbox{if $p$ and $q$ share a common edge};\\
            0 & \mbox{otherwise}\, ,
	   \end{array}
  \right. 
\end{equation}
where $\mu[p,q]=(-1)^{i+j}$ when 
the mid-point of the common edge 
corresponds to $x_i(p)$ as well as $x_j(q)$. 
We also note that the anticommutation relation 
\begin{equation}
\label{eq:anticom ab}
 \acom{a_{\alpha,\sigma}^\dagger,b_{p,\tau}} = 0
\end{equation}
holds for any $\alpha\in \calV$, any $p\in \calP$ and $\sigma,\tau=\up,\dn$. 
Then we define the hopping amplitudes $t_{x,y}$ so 
that the hopping term $\Hhop$ can be
written as
\begin{equation}
 \Hhop = -s \sumsigma \sum_{\alpha\in \calV} \aasd\aas 
         + t \sumsigma\sum_{p\in \calP} \bpsd\bps  
\end{equation}
with parameters $s\ge0$ and $t>0$.
Explicitly the hopping amplitudes are given by 
\begin{equation}
\label{eq:hopping matrix}
 t_{x,y} = \left\{\begin{array}{ll}
	         -2s+2(\nu-1)\,t & \mbox{ if $x=y$}; \\
                 -s    & \mbox{ if $|x-y|=1$ and 
                          $m(x,y)\in \calV$};\\ 
                 t    & \mbox{ if $|x-y|=1$ and
                          $m(x,y)\notin \calV$}; \\ 
                -(s+t)& \mbox{ if $|x-y|=\frac{\sqrt{2}}{2}$};\\
                 0    & \mbox{ otherwise},
		 \end{array}   
          \right.
\end{equation}
where $m(x,y)$ is defined to 
be the point taken in the middle of sites $x$ and $y$ in $\La$,
as in the case of $m(\alpha,\beta)$ for
$\alpha,\beta\in \calV$ (see Fig.~\ref{fig:lattice2d}). 

The main result of this paper is the occurrence of
ferromagnetism in the model defined above
at zero temperature:
\begin{theorem}
\label{th:main}
 Consider the Hubbard Hamiltonian $H$ with the
 hopping matrix given by~\eqref{eq:hopping matrix},
 and 
 suppose that the number $\Ne$ of electrons is $|\calV|$.
 Then, for each value of $s$ 
 there exist $t_\mathrm{c}$ and $U_\mathrm{c}$ 
 which are independent of the lattice size $|\calV|$
 such that for $t>t_\mathrm{c}$ and $U>U_\mathrm{c}$, 
 the following are valid:
 \begin{enumerate}
 \item the ground state energy is $-2\nu s|\calV|$,
 \item the ground states have the maximal total spin $\Stot = \Ne/2$,
 \item the ground state is unique apart {from} $(2\Stot+1)$-fold 
       degeneracy
       due to the spin rotation symmetry.
 \end{enumerate}
\end{theorem}

In the case $s=0$, 
the single-electron ground state energy is zero and
$|\calV|$-fold degenerate, as we will see in the next section.  
In this case, 
the model corresponds to 
the flat-band Hubbard model
and the ground state exhibits saturated ferromagnetism 
for all positive values of $t$ and $U$ 
($t_\mathrm{c}=0,U_\mathrm{c}=0$).
In particular, 
in the case $\nu=2$, our model with $s=0$ 
is equivalent to 
the flat-band Hubbard model on the line graph 
of the square lattice%
\footnote{%
One finds that our model with $s=0$ is unitary equivalent to
Mielke's flat-band model 
by performing a gauge transformation $c_{x,\sigma}\to -c_{x,\sigma}$
at all sites in one of the two sublattices, say, 
$\{x=m(\alpha,\alpha+\delta_1)|\alpha\in\calV\}$. 
},
and the occurrence
of saturated ferromagnetism in the model was proved 
by Mielke~\cite{Mielke91a,Mielke91b}.  
Our model in three or more dimensions, however,  
does not belong to the flat-band models on line graphs.
It is also noted that our model does not belong to 
the class of flat-band models discussed in detail 
by Mielke and Tasaki~\cite{MielkeTasaki93}.
We will give the proof of Theorem~\ref{th:main} for $s=0$ 
in Appendix~\ref{s:flat-band ferro}.

On the other hand, the lowest band of 
the model is dispersive for $s>0$, and the proof for such cases 
is quite different {from} that for the flat-band case. 
Here we adopt the same idea which is used for proving the occurrence of ferromagnetism
in Tasaki's models~\cite{Tasaki95,Tasaki2003}.
We first decompose the Hamiltonian $H$ into local Hamiltonians, and investigate
their properties, in particular, conditions for states to attain
the local minimum energy in detail.
Then the ferromagnetic state is shown to attain the minimum energy 
simultaneously for all the local Hamiltonians 
for $t>t_\mathrm{c}$ and $U>U_\mathrm{c}$.  
That is, the model is ``frustration free'' 
when the lower band supporting the ferromagnetism is almost flat 
and the on-site repulsion is sufficiently large. 
It is noted that, compared with Tasaki's nearly-flat band case, 
we need more lengthy discussion about 
local properties because of the gapless nature of our model.
We will devote ourselves to such discussion 
in section~\ref{s:local properties} 
and complete the proof for the case $s>0$ 
in section~\ref{s:proof of theorem}.
\section{Single-electron dispersion relations 
and basis states of the $\Ne$-electron space}
\label{s:single electron}
Let us consider the single-electron 
Schr\"odinger equation
for our Hubbard Hamiltonian.
Let $\Phi_{1,\sigma}$ be a single-electron state
\begin{equation}
 \Phi_{1,\sigma}=\sum_{x\in\La} \phi_x \cxsd \Phi_0\,,
\end{equation}
where $\phi_x$ is a complex coefficient 
and $\Phi_0$ is the state with no electron in $\La$.
Applying $H$ to $\Phi_{1,\sigma}$, we obtain 
the eigenvalue equation 
\begin{equation}
 \sum_{y\in\La}t_{x,y} \phi_y = \varepsilon \phi_x\,,
\end{equation} 
where $\varepsilon$ denotes a single-electron energy eigenvalue.

Because of the translation invariance of the Hamiltonian,
we can write an eigenstate $(\phi_x)_{x\in\La}$
in the form of the Bloch state as
$\phi_x = \rme^{\rmi k\cdot x}v_x(k)$,
where $v_x(k)$ satisfies $v_x(k)=v_{x+z}(k)$ for 
any $z\in\Bbb{Z}^d$ and $k$ is an element in $\calK$ defined by 
\begin{equation}
 \calK= 
 \left\{k=(k_1,\dots,k_\nu)~|~
 k_l = \frac{2\pi}{L}n_l,~n_l=0,\pm1,\dots, \pm\frac{L-1}{2}
 \mbox{ for $l=1,\dots,\nu$}
 \right\}\,.
\end{equation}
Then the eigenvalue equation is reduced to
\begin{equation}
\label{eq:eigenvalue equation}
\left(
-4s\cos^2\frac{k_l}{2}
+4t\sumtwo{m=1}{m\ne l}^\nu \cos^2\frac{k_m}{2}
\right) v_l
-4(s+t)\cos\frac{k_l}{2} 
 \sumtwo{m=1}{m\ne l}^\nu 
\left(\cos\frac{k_m}{2}
\right) v_m = \varepsilon v_l
\end{equation}
with $l=1,\dots,\nu$, where we write $v_l$ 
for $v_{{\delta_l}/{2}}(k)$ for notational simplicity.   

\begin{figure}
\begin{center}
 \includegraphics[scale=1]{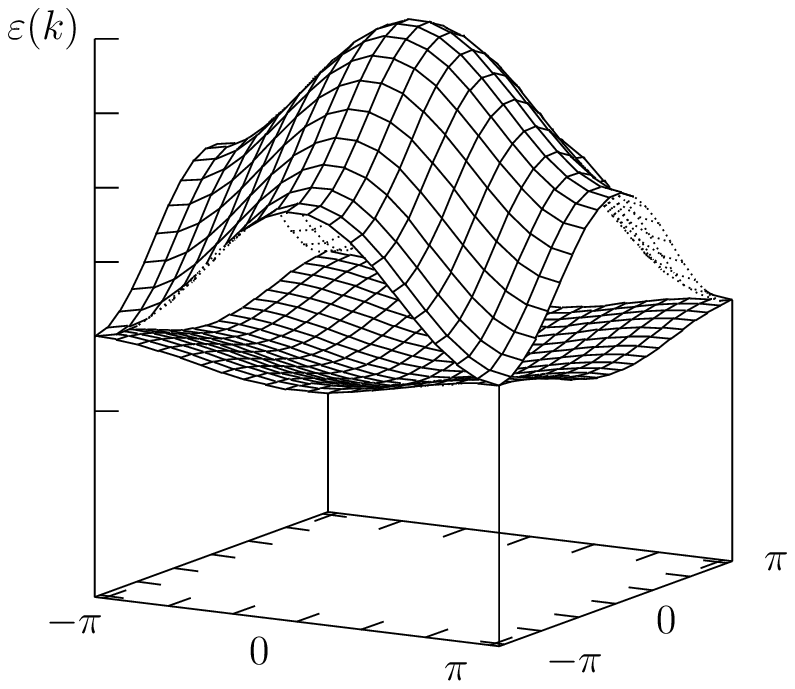}
\end{center}
\caption{The dispersion relations for $\nu=2$.}
\label{fig:dispersion}
\end{figure}

Setting $\mathsf{A}_{l,m} = \cos(k_l/2)\cos(k_m/2)$
and 
\begin{equation}
 \varepsilon^\prime = 
-\frac{1}{s+t}\left(
              \frac{\varepsilon}{4}-t\sum_{l=1}^\nu \cos^2 \frac{k_l}{2}
              \right)\,,
\end{equation}
we can further rewrite \eqref{eq:eigenvalue equation} as
\begin{equation}
 \sum_{m=1}^\nu \mathsf{A}_{l,m} v_m = \varepsilon^\prime v_l\,.
\end{equation}
It is easy to see that the rank of matrix 
$\mathsf{A}=[\mathsf{A}_{l,m}]_{1\le l,m\le \nu}$
is $1$, and therefore $\varepsilon^\prime=0$ is the $(\nu-1)$-fold degenerate
eigenvalue of $\mathsf{A}$.
It is also easy to see that the vector $(v_m)_{m=1}^{\nu}=(\cos(k_1/2),\cos(k_2/2),\dots,\cos(k_\nu/2))$ 
is an eigenstate of $\mathsf{A}$ and
its eigenvalue is given by
\begin{equation}
 \varepsilon^\prime=\sum_{l=1}^\nu \cos^2 \frac{k_l}{2}\,.
\end{equation}
As a result we obtain the single-electron energy eigenvalues
which are characterized by the dispersion relations 
\begin{equation}
\varepsilon_0(k) =  -4s \sum_{l=1}^\nu \cos^2\frac{k_l}{2}
                 =  -2s \sum_{l=1}^\nu (1+\cos k_l)
\end{equation}
and
\begin{equation}
\varepsilon_1(k) =  4t \sum_{l=1}^\nu \cos^2\frac{k_l}{2}
                 =  2t \sum_{l=1}^\nu (1+\cos k_l)\,.
\end{equation}
The eigenvalue $\varepsilon_1(k)$ 
is $(\nu-1)$-fold degenerate for each $k$.  
See Fig.~\ref{fig:dispersion} for the dispersion relations
for $\nu=2$.

In the case of $s=0$, $\varepsilon_0(k)=0$ for all $k\in \calK$,
i.e., the single-electron ground state energy is zero and
$|\calV|$-fold degenerate.

In the rest of this section we comment 
on a construction of $\Ne$-electron states.

Let us denote by $f_{(0,k),\sigma}$ 
a fermion operator corresponding to the eigenstate 
with the eigenvalue $\varepsilon_0(k)$.
We also denote by $f_{(l,k),\sigma}$  with $l=1,\dots,\nu-1$  
fermion operators corresponding to linearly independent
$\nu-1$ eigenstates with the eigenvalue $\varepsilon_1(k)$.
Then, owing to linear independence of the energy eigenstates, 
any $\Ne$-electron state can be
represented as a linear combination of the states
\begin{equation}
 \left(\prod_{i\in A_\up} f_{i,\up}^\dagger\right) 
 \left(\prod_{i\in A_\dn} f_{i,\dn}^\dagger\right)\Phi_0
\end{equation}
with $|A_\up|+|A_\dn|=\Ne$, where $A_\up$ and $A_\dn$
are arbitrary subsets of
\begin{equation}
 \{i=(l,k)~|~l=0,\dots,\nu-1,k\in \calK\}\,.
\end{equation}

It is crucial to find that the fermion operators $f_{(0,k),\sigma}$
can be expanded as a linear combination of $\aas$.
This is observed as follows.
First we note that
$\sum_{\alpha\in \calV} u_\alpha \aasd\Phi_0$
is zero if and only if $u_\alpha = 0$ 
for all $\alpha\in \calV$, i.e.,
the states $\aasd\Phi_0$ are linearly independent.
Next consider the single-electron   
Schr\"odinger equation
\begin{equation}
 H \sum_{\beta\in \calV}\phi_\beta a_{\beta,\sigma}^\dagger \Phi_0
 = \varepsilon \sum_{\beta\in \calV}
   \phi_\beta a_{\beta,\sigma}^\dagger\Phi_0\,, 
\end{equation}
where $\phi_\beta$ are complex coefficients
and $\varepsilon$ is real.
By using the anticommutation relations \eqref{eq:anticom aa} and
\eqref{eq:anticom ab}, we reduce the left-hand side of the above
equation to a linear combination of $a_{\beta,\sigma}^\dagger\Phi_0$.
Then, comparing the coefficients of $a_{\beta,\sigma}^\dagger\Phi_0$ 
in the left- and the right-hand sides, we obtain
\begin{equation}
 -2\nu s \phi_{\beta} 
 -s\sumtwo{\alpha\in \calV}{|\alpha-\beta|=1} 
                       \phi_\alpha
 = \varepsilon \phi_\beta \,.
\end{equation}
By solving the above equation we obtain 
the eigenvalue $\varepsilon_0(k)$ with $k\in \calK$.
Therefore the sets 
$\{f_{(0,k),\sigma}^\dagger\Phi_0\}_{k\in \calK}$
and $\{a_{\alpha,\sigma}^\dagger\Phi_0\}_{\alpha\in \calV}$ 
span the same space.   
This also indicates the fact that 
any $\Ne$-electron state can be
represented as a linear combination of the states
\begin{equation}
\label{eq:basis}
\Phi(V_\up,B_\up;V_\dn,B_\dn)
=
 \left(\prod_{\alpha\in V_\up} a_{\alpha,\up}^\dagger\right) 
 \left(\prod_{i\in B_\up} f_{i,\up}^\dagger\right) 
 \left(\prod_{\alpha\in V_\dn} a_{\alpha,\dn}^\dagger\right)
 \left(\prod_{i\in B_\dn} f_{i,\dn}^\dagger\right)\Phi_0
\end{equation}
with $|V_\up|+|V_\dn|+|B_\up|+|B_\dn|=\Ne$, 
where $V_\up$ and $V_\dn$ are arbitrary subsets of $\calV$,
and 
$B_\up$ and $B_\dn$
are arbitrary subsets of
\begin{equation}
 \{i=(l,k)~|~l=1,\dots,\nu-1,k\in \calK\}\,.
\end{equation}

\section{Local properties of the model}
\label{s:local properties}
In this section we represent
the Hamiltonian \eqref{eq:Hamiltonian}
as a sum of local Hamiltonians
and investigate properties of the local Hamiltonians.
As a consequence we will obtain a lemma which will 
play a central role in the proof of
Theorem~\ref{th:main} for $s>0$.

\subsection{Decomposition of the Hamiltonian}
For each vertex $\alpha\in \calV$, 
we define the sublattice $\Laa$ as
\begin{equation}
 \Laa = \{ x=x_i(p)~|~i=1,\dots,4,~p\in \calP_\alpha\}\,,
\end{equation}
which is the collection of the sites on the edges of the
regular squares which share vertex $\alpha$.
(In the case $\nu=2$, $\Laa$ consists of 12 filled circles
depicted in Fig.\ref{fig:lattice2d}.)
Then we define the local Hamiltonian $\halph$ on the lattice $\Laa$ by
\begin{equation}
\label{eq:local Hamiltonian}
 \halph  =   
        -s\sumsigma\aasd\aas 
          + \frac{t}{4}\sumsigma\sum_{p\in \calP_\alpha}
              b_{p,\sigma}^\dagger b_{p,\sigma}
          +\frac{U}{2(2\nu-1)}\sum_{x\in\Laa} \nxup\nxdn\, .
\end{equation}
By using $\halph$, the Hamiltonian $H$ is decomposed as
\begin{equation}
 H = \sum_{\alpha\in \calV} \halph\, .
\end{equation} 
It is noted that the local Hamiltonians do not commute with one another,
and it is thus not trivial to find out something about 
the whole system, such as the energy and  magnetic
properties of the ground state, by using these local Hamiltonians. 

The properties of the local Hamiltonian $\halph$ 
for sufficiently large values of $U/s$ and $t/s$ are 
summarized in the following lemma.
\begin{lemma}
\label{lemma:main}
Consider the local Hamiltonian $h_\alpha$
on the Hilbert space where
the electron number on the local lattice $\La_\alpha$ 
is not fixed.
Suppose that $t/s$ and $U/s$ are sufficiently large.
Then the minimum eigenvalue of $\halph$ is $-2\nu s$ and
any eigenstate $\Phi$ with this eigenvalue can be expressed as
\begin{equation}
\label{eq:condition1 in Lemma}
 \Phi = a_{\alpha,\up}^\dagger\Phi_\up + a_{\alpha,\dn}^\dagger\Phi_\dn, 
\end{equation} 
by using some states $\Phi_\up$ with $a_{\alpha,\dn}\Phi_\up=0$
and $\Phi_\dn$ with $a_{\alpha,\up}\Phi_\dn=0$.
Furthermore, $\Phi$ satisfies 
\begin{equation}
\label{eq:condition2 in Lemma}
 c_{x,\dn}c_{x,\up}\Phi = 0
\end{equation}
for all $x\in\La_\alpha$. 
\end{lemma}
The above lemma claims that electrons do not 
doubly occupy the localized states corresponding 
to $\aas$
when the strength of the repulsive interaction is 
sufficiently large and 
the high-energy part of local single-electron energy eigenvalues 
is well separated from the lower levels.

It is remarked that a similar lemma is proved 
for Tasaki's models and 
the proof of our lemma proceeds in almost the same way as 
in Refs.~\cite{Tasaki95,Tasaki2003}.
However, the proof for our case is much more involved.
The difference is due to the following reason.
The localized states corresponding to $a$-operators in Tasaki's models 
contain a parameter which controls a band gap or 
hopping amplitudes.
By taking a certain limit of this parameter,
the local Hamiltonians of Tasaki's models are
reduced to ``atomic Hamiltonians'', which have no hoppings of electrons.
This fact makes it somewhat easier to treat the local Hamiltonians.
On the other hand, our model does not contain such a parameter, and
we have to prove the lemma by directly using eigenstates of~$h_\alpha$. 

It is also remarked that the local Hamiltonian $h_\alpha$ is supported on 
the sublattice consisting of $2\nu(2\nu-1)$ sites, 
which becomes a 12-site lattice even in two dimensions, and
the decomposition with smaller sublattices probably does not work. 
It is a non-trivial task to analytically solve a problem of interacting electrons 
on this lattice size%
\footnote{%
In Ref.~\cite{Lu2009}, a similar lemma is proved relying on numerical calculations.
}.   
\subsection{Single-electron problem for the local Hamiltonian}
Because of the translation invariance, it suffices to
investigate $h_0$, the local Hamiltonian associated with the origin $0=(0,\dots,0)$
of $\calV$, in order to prove Lemma~\ref{lemma:main}.
Since $h_0$ is defined only on $\Lambda_0$,
we consider the space constructed by using
fermion operators   
$c_{x,\sigma}$ with $x\in\La_0$, $\sigma=\up,\dn$.

In this subsection 
we will solve the single-electron problem for $h_0$.
As usual we can solve it
by using basis states $\{\cxsd\Phi_0\}_{x\in\La_0}$
of the single-electron Hilbert space $\calH$ on $\La_0$, 
but we adopt a slightly different method here.

To prove Lemma~\ref{lemma:main}, 
we will consider finite energy states 
in the limit $t\to\infty$.
So let us firstly characterize single-electron states
which have infinitely large energy in the limit $t\to\infty$.
Let $\calH_b$ be the single-electron Hilbert space spanned by the states
$\bpsd\Phi_0$ with $p\in \calP_0$, which are linearly independent.
Since all the creation operators $\bpsd$ 
anticommute with $a_{0,\sigma}$,
we find that 
$h_0 \Phi_{1,\sigma}\in \calH_b$ 
for any $\Phi_{1,\sigma}\in \calH_b$.
This implies that it is possible to obtain eigenstates of $h_0$ 
within $\calH_b$.
Since the operator $t\sum_{p\in \calP_0}\bpsd\bps$ 
restricted on $\calH_b$
is positive definite,
corresponding energy eigenvalues are proportional to $t$,
and thus $\calH_b$ consists of high energy states.

Let $\calH_b^{\perp}$ be the orthogonal complement 
of $\calH_b$ within $\calH$.
In the following, we solve the single-electron problem for $h_0$
within $\calH_b^{\perp}$, by constructing states orthogonal to
those in $\calH_b$ 
(or equivalently 
by constructing local fermion operators anticommuting 
with $\bpsd$).    

Let $\dLao$ be the collection of the ``boundary'' sites 
in $\Lambda_0$
\begin{equation}
 \dLao = \Lambda_0 \backslash \calC_0\,.
 \label{eq:boundary}
\end{equation}
Without loss of generality, we can assume 
that ${x}_3(p),{x}_4(p)\in\dLao$ for all $p\in \calP_0$.
Then, we define 
\begin{equation}
\label{eq:dps}
 \dps = \frac{1}{\sqrt{2}}(c_{{x}_3(p)} + c_{{x}_4(p)}) 
\end{equation}
for each $p\in \calP_0$ and  
\begin{equation}
\label{eq:bdxs}
 \bar{d}_{x,\sigma} = 2 \cxs + \sum_{p\in \calP(x)}
                      \mu[x,p](c_{x_3(p),\sigma} - c_{x_4(p),\sigma})\, , 
\end{equation}
for each $x\in \calC_0$ ,
where $\mu[x,p]$ takes $-1$ if $x$ corresponds to $x_1(p)$ 
and 1 if $x$ corresponds to $x_2(p)$.
By a straightforward calculation, we can check that
\begin{equation}
 \acom{b_{p,\sigma}^\dagger, d_{q,\tau}} =
 \acom{b_{p,\sigma}^\dagger, \bar{d}_{x,\tau}} = 0
\end{equation}
for any $p,q\in \calP_0$ and $x\in \calC_0$. 
It is also not difficult to check that
$|\calP_0|+|\calC_0|$ states, ${d}_{p,\sigma}^\dagger\Phi_0$ with $p\in \calP_0$
and $\bar{d}_{x,\sigma}^\dagger\Phi_0$ with $x\in \calC_0$, are
linearly independent.
Noting that the dimension of $\calH_b$ is $|\calP_0|$,
and $|\calP_0|+(|\calP_0|+|\calC_0|) = |\La_0|$,
we find that
these states   
span the orthogonal complement $\calH_b^\perp$ of $\calH_b$.

Since the operator~$s a_{0,\sigma}^\dagger a_{0,\sigma}$ 
restricted to $\calH$
is a projection onto the space spanned by 
the single state $a_{0,\sigma}^\dagger\Phi_0$,
$h_0$ restricted to $\calH_b^\perp$
has exactly two eigenvalue; one is 0 and 
the other is $-2\nu s$ to which $a_{0,\sigma}^\dagger\Phi_0$ 
belongs.
 
One easily finds that all $\dpsd$ with $p\in\calP_0$,
which are supported only on the sites in $\partial\La_0$,
anticommute 
with not only $\bps$ but also $a_{0,\sigma}$,
so that the states $\dpsd\Phi$ 
are the eigenstates of $h_0$
with eigenvalue 0.
Furthermore, it follows {from} $\acom{\dpsd,d_{q,\sigma}}=\delta_{p,q}$ 
that the set of states $\left\{\dpsd\Phi_0\right\}_{p\in\calP_0}$ 
is orthonormal.
   
Let us determine the rest of eigenstates with eigenvalue 0.
Since the anticommutation relation
\begin{equation}
 \acom{d_{p,\sigma}^\dagger, \bar{d}_{x,\sigma}} = 0
\end{equation}
holds for $p\in \calP_0$ and $x\in \calC_0$,
our task is to construct fermion operators
which anticommute with $a_{0,\sigma}^\dagger$,
by using $\bar{d}_{x,\sigma}$
with $x\in \calC_0$. 
In the following, we write $x_l$ for $m(0,\delta_l)$
and $x_{-l}$ for $m(0,-\delta_l)$ with $l=1,\dots \nu$.  
We note that
\begin{equation}
 \calC_0 = \{ x_l~|~l=\pm1,\dots,\pm \nu\}\,.
\end{equation}

By using $\bar{d}_{x,\sigma}$, we form fermion operators which anticommute
with one another.
We write again $d_{i,\sigma}$ for these fermion operators.
To label new $d$-operators which we will define below,
we introduce a set $\calD=\left\{1,\dots,\nu\right\}$, and also 
\begin{equation}
\calJ=\left\{0,\pm\frac{2\pi}{\nu},\dots,\pm\frac{2\pi}{\nu}\frac{(\nu-1)}{2} \right\} 
\end{equation}
for odd $\nu$,
and 
\begin{equation}
\calJ=\left\{0,\pm\frac{2\pi}{\nu},\dots,\pm\frac{2\pi}{\nu}\frac{\nu-2}{2},\pi \right\} 
\end{equation}
for even $\nu$.

Now, for $l\in \calD$, let us
define
\begin{equation}
\label{eq:dls}
d_{l,\sigma} = \frac{1}{2\sqrt{2\nu}}
         (\bar{d}_{x_l,\sigma}-\bar{d}_{x_{-l},\sigma})\,.    
\end{equation}
We also define
\begin{equation}
\label{eq:dks}
 d_{k,\sigma} = \frac{1}{2\sqrt{2\nu(\nu+1-\nu\delta_{k,0})}}
                \sum_{l=1}^\nu
		\rme^{\rmi kl} (\bar{d}_{x_l,\sigma}
                               +\bar{d}_{x_{-l},\sigma})
\end{equation}
for $k\in\calJ$.
We note that $d_{k,\sigma}$ with $k=0$ is
equal to $(1/\sqrt{2\nu})a_{0,\sigma}$.
Furthermore, as we will see below, we have the anticommutation relations
\begin{equation}
 \acom{d_{i,\sigma}^\dagger, d_{j,\sigma}} = \delta_{i,j}
 \label{eq:anticom}
\end{equation}
for any $i,j\in \calD\cup \calJ$.
Therefore, the single-electron states  
$d_{i,\sigma}^\dagger\Phi_0$ with $i\in \calD\cup\calJ\backslash\{0\}$
in $\calH_b^\perp$  
are orthonormal eigenstates of $h_0$ with eigenvalue 0.

To prove~\eqref{eq:anticom}, let us first calculate the anticommutation
relations for $\bar{d}_{x,\sigma}^\dagger$
and $\bar{d}_{y,\sigma}$ with $x,y\in \calC_0$.
Noting that $|\calP(x_l)|=2(\nu-1)$ and 
$|\calP(x_l)\cap \calP(x_{-l})|=\emptyset$ 
for $l=\pm1,\dots,\pm \nu$, 
we find 
\begin{equation}
\label{eq:anticom td1}
 \acom{\bar{d}_{x_l,\sigma}^\dagger, 
                \bar{d}_{x_l,\sigma}}=4+2|\calP(x_l)|=4\nu\,,  
\end{equation}
and  
\begin{equation}
\label{eq:anticom td2}
 \acom{\bar{d}_{x_l,\sigma}^\dagger, \bar{d}_{x_{-l},\sigma}} = 0\, .
\end{equation}
Next, noting that, for 
$l,l^\prime=\pm1,\dots,\pm \nu$ and $l\ne \pm l^\prime$,
$\calP(x_l)\cap \calP(x_{l^\prime})$ always has
only one element, say $p$, such that
either $x_l=x_1(p),x_{l^\prime}=x_2(p)$ or 
$x_l=x_2(p),x_{l^\prime}=x_1(p)$
holds%
\footnote{
The regular square $p$ is in the plane including the $l$-axis and 
the $l^\prime$-axis.
},
we have
\begin{equation}
\label{eq:anticom td3}
 \acom{\bar{d}_{x_l,\sigma}^\dagger, 
       \bar{d}_{x_{l^\prime},\sigma}} = 2\mu[x_l,p]\mu[x_{l^\prime},p]
                                      =-2.
\end{equation}
Then, by using anticommutation 
relations~\eqref{eq:anticom td1}, 
         \eqref{eq:anticom td2} and 
         \eqref{eq:anticom td3},
we obtain
\begin{equation}
\label{eq:anticom td4}
 \{(\bar{d}_{x_l,\sigma}^\dagger-\bar{d}_{x_{-l},\sigma}^\dagger), 
       (\bar{d}_{x_{l^\prime},\sigma}-\bar{d}_{x_{-l^\prime},\sigma})\}
 ={8\nu}\delta_{l,l^\prime},
\end{equation}
\begin{equation}
\label{eq:anticom td5}
 \{(\bar{d}_{x_l,\sigma}^\dagger+\bar{d}_{x_{-l},\sigma}^\dagger), 
       (\bar{d}_{x_{l^\prime},\sigma}+\bar{d}_{x_{-l^\prime},\sigma})\}
 ={8\nu}\delta_{l,l^\prime}-8(1-\delta_{l,l^\prime}),
\end{equation}
and
\begin{equation}
\label{eq:anticom td6}
 \{(\bar{d}_{x_l,\sigma}^\dagger+\bar{d}_{x_{-l},\sigma}^\dagger), 
       (\bar{d}_{x_{l^\prime},\sigma}-\bar{d}_{x_{-l^\prime},\sigma})\}
 =0
\end{equation}
for $l,l^\prime\in\calD$.
From, \eqref{eq:anticom td4}, 
         \eqref{eq:anticom td5} and 
         \eqref{eq:anticom td6},
we obtain the desired relations~\eqref{eq:anticom}.

The results in this subsection are summarized 
as follows.
Let $\calI=\calP_0\cup \calD \cup \calJ$.
The fermion operators defined by \eqref{eq:dps}, \eqref{eq:dls} and \eqref{eq:dks} 
satisfy
the anticommutation relation
\begin{equation}
 \acom{d_{i,\sigma}^\dagger, d_{j,\sigma}} = \delta_{i,j}
\end{equation}   
for any $i,j\in \calI$.
The single-electron states $d_{i,\sigma}^\dagger\Phi_0$ 
with $i\in \calI$ 
satisfy 
\begin{equation}
h_0 d_{i,\sigma}^\dagger\Phi_0 
        = \varepsilon_i d_{i,\sigma}^\dagger\Phi_0\,,
\end{equation}
where $\varepsilon_i$ are given by
\begin{equation}
 \varepsilon_i = \left\{\begin{array}{ll}
              -2\nu s & \mbox{if $i=0$};\\
               0   & \mbox{otherwise}, 
		\end{array}
               \right.
\end{equation}
and they form an orthonormal basis%
\footnote{%
Note that $|\calI|=|\calP_0|+|\calD|+|\calJ|=|\calP_0|+|\calC_0|$ which equals the dimension of $\calH_b^\perp$.
}
for subspace $\calH_b^\perp$.
\subsection{Proof of Lemma~\ref{lemma:main}}
By using the local fermion operators introduced in the previous section,
we solve a many-electron problem for $h_0$.
First, we will consider the problem in the limit $t,U\to\infty$, 
where it will be proved that the minimum expectation value of $h_0$
is equal to or greater than $-2\nu s$ 
and that any state which attains the minimum expectation value
is an eigenstate of $h_0$. 

Let $\Phi$ be a state on $\La_0$ with a finite energy 
in the limit $t,U\to\infty$.
Consider representing $\Phi$ by  
using the fermion operators 
$\{d_{i,\sigma}^\dagger\}_{i\in \calI}$
and 
$\{b_{p,\sigma}^\dagger\}_{p\in \calP_0}$.
In the limit $t\to\infty$, since the terms 
$t b_{p,\sigma}^\dagger b_{p,\sigma}$ in $h_0$ are positive semidefinite,
$\Phi$ with a finite energy must satisfy 
$b_{p,\sigma}\Phi=0$ for all $p\in \calP_0$ and
$\sigma=\up,\dn$.
This means that a finite energy state $\Phi$ is written as   
\begin{equation}
\label{eq:finite energy state}
 \Phi = \sum_{I_\up,I_\dn\subset \calI}
          g(I_\up;I_\dn) \Phi(I_\up,I_\dn)
\end{equation}
with complex coefficients $g(I_\up;I_\dn)$,
where 
\begin{equation}
\label{eq:d basis}
  \Phi(I_\up;I_\dn) = \left(\prod_{i\in I_\up}d_{i,\up}^\dagger\right)  
          \left(\prod_{j\in I_\dn}d_{j,\dn}^\dagger\right)\Phi_0 \,.
\end{equation}

Here, to avoid an ambiguity which may arise due to 
the exchange of fermion operators, 
we adopt the following rule for the product
of fermion operators.
Let $\theta$ be a one-to-one mapping {from} $\calI$ to $\Bbb{Z}$.
Then, we assume that the products of 
the fermion operators in $\Phi(I_\up;I_\dn)$ are ordered
in such a way that   
$d_{i,\up}^\dagger$ (respectively, $d_{i,\dn}^\dagger$)
is always on the left of 
$d_{j,\up}^\dagger$ (respectively, $d_{j,\dn}^\dagger$)
if $\theta(i)<\theta(j)$.
For later use, we also define  
\begin{equation}
 \label{eq:sign factor S}
  \mathsf{S}_{I_\sigma}^{i} 
  =\prod_{j\in I_{\sigma};\theta(j)<\theta(i)}(-1).
\end{equation}
For example, we have
\begin{equation}
 \left(\prod_{j\in I_{\sigma}} d_{j,\sigma}^\dagger\right)
 = 
 \mathsf{S}_{I_\sigma}^{i} 
 d_{i,\sigma}^\dagger 
 \left(\prod_{j\in I_{\sigma}\backslash \{i\}} d_{j,\sigma}^\dagger\right).
\end{equation}
 
Let us consider furthermore the limit $U\to\infty$.
Since the on-site interaction 
$U\nxup\nxdn=Uc_{x,\up}^\dagger c_{x,\dn}^\dagger c_{x,\dn}c_{x,\up}$
is positive semidefinite,
a state $\Phi$ with finite energy satisfies 
$c_{x,\dn}c_{x,\up}\Phi=0$ for any $x\in\La_0$.  
Substituting \eqref{eq:finite energy state}
into this equation, we find that the coefficients 
$g(I_\up;I_\dn)$ must be chosen so that 
\begin{equation}
 \sum_{I_\up,I_\dn\subset \calI}
          g(I_\up;I_\dn)
          c_{x,\dn}c_{x,\up}\Phi(I_\up;I_\dn)= 0\,
\label{eq:finite energy condition on g}
\end{equation}
will always hold for any $x\in\La_0$.

By using the coefficients $g(I_\up;I_\dn)$ 
which satisfy the condition~\eqref{eq:finite energy condition on g},
let us now express
an expectation value $E[\Phi]$ of $h_0$ for $\Phi$ 
in the limit of $t,U\to\infty$.
Since the eigenvalues of 
$-s \sum_{\sigma} a_{0,\sigma}^\dagger a_{0,\sigma}$
for $\Phi(I_\up;I_\dn)$  
are simply given by
\begin{equation}
 \left\{ \begin{array}{ll}
          -4\nu s & \mbox{if $0\in I_\up\cap I_\dn$}; \\
	  -2\nu s & \mbox{if $0\in I_\up\cup I_\dn$ and $0\notin I_\up\cap I_\dn$}; \\
           0   & \mbox{otherwise},
	 \end{array}
\right.
\end{equation}
and the set of all states $\Phi(I_\up;I_\dn)$ is orthonormal,
we have
\begin{eqnarray}
 E[\Phi] & = & \frac{\langle \Phi, h_0 \Phi \rangle}
                    {\langle \Phi, \Phi \rangle} \\
         & = & \left(-4\nu s \sumtwo{I_\up,I_\dn\subset \calI}
                                 {0\in I_\up\cap I_\dn} |g(I_\up;I_\dn)|^2
                     -2\nu s \sumtwo {I_\up,I_\dn\subset \calI}
                                 {0\in I_\up\cup I_\dn, 0\notin I_\up\cap I_\dn}
                                                        |g(I_\up;I_\dn)|^2 
               \right)||\Phi||^{-2}
\end{eqnarray}
with 
\begin{equation}
 ||\Phi||=\sqrt{\langle \Phi, \Phi \rangle}=\left(\sumIud|g(I_\up;I_\dn)|^2 \right)^\frac{1}{2}.
\end{equation}
By noting that
\begin{equation}
  \sumtwo{I_\up,I_\dn\subset \calI}{0\in I_\up\cap I_\dn}
  +\sumtwo{I_\up,I_\dn\subset \calI}{0\in I_\up\cup I_\dn, 0\notin I_\up\cap I_\dn}
  +\sumtwo{I_\up,I_\dn\subset \calI}{0\notin I_\up\cup I_\dn}
 =
  \sum_{I_\up,I_\dn\subset \calI},
\end{equation}
we rewrite $E[\Phi]$ as
\begin{equation}
 E[\Phi]=-2\nu s + 2\nu s F[\Phi]||\Phi||^{-2}
\end{equation}
where
\begin{eqnarray}
 F[\Phi]
  &=& \sumtwo{I_\up,I_\dn\subset \calI}{0\notin I_\up\cup I_\dn}|g(I_\up;I_\dn)|^2
      -\sumtwo{I_\up,I_\dn\subset \calI}{0\in I_\up\cap I_\dn}|g(I_\up;I_\dn)|^2
  \nonumber\\
  &=& \sumIudo |g(I_\up;I_\dn)|^2
          -\sumIudo |\g{0}{0}|^2 \,.
\end{eqnarray}
In the following, we show $F[\Phi]\ge0$. This implies $E[\Phi]\ge -2\nu s$ since $s>0$.

In order to prove $F[\Phi]\ge 0$ we investigate in detail the conditions on 
$g(I_\up;I_\dn)$ imposed by~\eqref{eq:finite energy condition on g}.
The left-hand side of~\eqref{eq:finite energy condition on g} is calculated as
\begin{equation}
\label{eq:ccPhi}
 \sumtwo{I_\up,I_\dn\subset \calI}{|I_\up|\ge1,|I_\dn|\ge1}
          g(I_\up;I_\dn)
          \sum_{n\in I_\up}\sum_{m\in I_\dn}
	  (-1)^{|I_\up|-1}{\mathsf{S}_{I_\up}^n}{\mathsf{S}_{I_\dn}^m}
	  (\varphi_x^{(n)})^\ast(\varphi_x^{(m)})^\ast
          \Phi(I_\up\backslash\{n\};I_\dn\backslash\{m\})\, , 
\end{equation}
where 
$(\varphi_x^{(i)})^\ast=\{d_{i,\sigma}^\dagger,c_{x,\sigma}\}$.
The expression \eqref{eq:ccPhi} is further calculated as
\begin{eqnarray}
\label{eq:ccPhi2}
\lefteqn{\sum_{n,m\in \calI}
  (\varphi_x^{(n)})^\ast(\varphi_x^{(m)})^\ast 
  \sumtwo{I_\up,I_\dn\subset \calI}{|I_\up|\ge1,|I_\dn|\ge1}
  (-1)^{|I_\up|-1}
  {\mathsf{S}_{I_\up}^n}{\mathsf{S}_{I_\dn}^m}
  \chi[n\in I_\up]\chi[m\in I_\dn]}\nonumber\\	  
&& \hspace*{7cm}\times
          g(I_\up;I_\dn)          
	  \Phi(I_\up\backslash\{n\};I_\dn\backslash\{m\})\nonumber\\ \nonumber\\
&=&
 \sum_{n,m\in \calI}
  (\varphi_x^{(n)})^\ast(\varphi_x^{(m)})^\ast 
  \sum_{I_\up\subset {\calI}\backslash\{n\}}
  \sum_{I_\dn\subset {\calI}\backslash\{m\}}
          (-1)^{|I_\up|}
	  {\mathsf{S}_{I_\up}^n}{\mathsf{S}_{I_\dn}^m}	  
          g(\{n\}\cup I_\up;\{m\}\cup I_\dn)          
          \Phi(I_\up;I_\dn) \nonumber\\
&=&
 \sumIud
 \sum_{n,m\in \calI}
  (\varphi_x^{(n)})^\ast(\varphi_x^{(m)})^\ast 
          \tg{n}{m}\Phi(I_\up;I_\dn)\, ,
\end{eqnarray}
where $\chi[\textrm{event}]$ is the indicator function
which takes 1 if event is true and 0 otherwise, and
we have introduced the subsidiary coefficients defined as
\begin{equation}
\label{eq:tg1}
 \tg{n}{m} = \left\{
	      \begin{array}{ll}
	       0 & \mbox{if $n\in I_\up$ or $m\in I_\dn$;}\\
	       (-1)^{|I_\up|}
		{\mathsf{S}_{I_\up}^n}{\mathsf{S}_{I_\dn}^m}
		\g{n}{m}
	       &\mbox{otherwise.}
	      \end{array}
       \right.
\end{equation}
In the first equality of \eqref{eq:ccPhi2} we have used 
${\mathsf{S}_{I_\sigma\cup\{i\}}^i}={\mathsf{S}_{I_\sigma}^i}$. 
Since all the states $\Phi(I_\up;I_\dn)$ are linearly independent,
we find that the condition~\eqref{eq:finite energy condition on g} 
becomes
\begin{equation}
\label{eq:finite energy condition}
  \sum_{n,m\in \calI}
  (\varphi_x^{(n)})^\ast(\varphi_x^{(m)})^\ast 
          \tg{n}{m} = 0
\end{equation}
for any $I_\up,I_\dn\subset {\calI}$ and for any $x\in\La_0$.

By choosing the sites in $\calC_0$ as $x$, 
we rewrite condition~\eqref{eq:finite energy condition}
more concretely.
Take $x_l=m(0,\delta_l)\in \calC_0$ as $x$ in condition~\eqref{eq:finite energy condition}.
Noting that $\varphi_{x_l}^{(i)}$ is vanishing if 
$i\in \calP_0$ or $i\in \calD\backslash\{l\}$%
\footnote{%
We calculate the coefficients $\varphi_{x_l}^{(i)}=\{d_{i,\sigma}^\dagger, c_{x_l}\}$ 
by using
\eqref{eq:dps}, \eqref{eq:bdxs}, \eqref{eq:dls} and \eqref{eq:dks}.
},
we obtain 
\begin{eqnarray}
\lefteqn{\tilde{g}(l;l) 
+ \tilde{g}(0;0)
+ \frac{1}{\nu+1}\sum_{k,k^\prime\in\calJ\backslash\{0\}}
                    \rme^{-\rmi kl}\rme^{-\rmi k^\prime l}
                    \tilde{g}(k;k^\prime)} \nonumber \\
&&
+ (\tilde{g}(l;0)+\tilde{g}(0;l))
+ \frac{1}{\sqrt{\nu+1}}
                   \sumJo
                    \rme^{-\rmi kl}(\tilde{g}(l;k)+\tilde{g}(k;l)) \nonumber\\
&&
+ \frac{1}{\sqrt{\nu+1}}
                   \sumJo
                    \rme^{-\rmi kl}(\tilde{g}(0;k)+\tilde{g}(k;0)) = 0\, ,              
\end{eqnarray}
where we write $\tilde{g}(n;m)$ for 
$\tg{n}{m}$.
Similarly, for $x_{-l}=m(0,-\delta_l)\in \calC_0$, we obtain
\begin{eqnarray}
\lefteqn{\tilde{g}(l;l) 
+ \tilde{g}(0;0)
+ \frac{1}{\nu+1}\sum_{k,k^\prime\in\calJ\backslash\{0\}}
                    \rme^{-\rmi kl}\rme^{-\rmi k^\prime l}
                    \tilde{g}(k;k^\prime)} \nonumber \\
&&
- (\tilde{g}(l;0)+\tilde{g}(0;l))
- \frac{1}{\sqrt{\nu+1}}
                   \sumJo
                    \rme^{-\rmi kl}(\tilde{g}(l;k)+\tilde{g}(k;l)) \nonumber\\
&&
+ \frac{1}{\sqrt{\nu+1}}
                   \sumJo
                    \rme^{-\rmi kl}(\tilde{g}(0;k)+\tilde{g}(k;0)) = 0\, .              
\end{eqnarray}
Then, summing the above two equations over $l\in \calD$ 
we obtain the condition%
\footnote{%
We think of $-\pi$ as $\pi$ when $\nu$ is even ($\pi$ is in $\calJ$).}
\begin{equation}
\label{eq:final condition}
 \tg{0}{0} =
- \frac{1}{\nu}\sum_{l\in \calD}\tg{l}{l} 
- \frac{1}{(\nu+1)}\sumJo \tg{k}{-k}\,.
\end{equation}
Our analysis below highly relies on the above relation.

Let $\bar{I}_\dn$ be the set defined by
\begin{equation}
 \bar{I}_\dn = \{ i ~|~i\in I_\dn\cap (\calP_0\cup D)\}
               \cup
	       \{ -i ~|~i\in I_\dn\cap \calJ\}\,,
\end{equation}
and let $N(I_\up;I_\dn)$ be the number of elements in
$I_\up\cap\bar{I}_\dn \cap (\calD\cup\calJ\backslash\{0\})$.
Condition~\eqref{eq:final condition} relates 
$\g{0}{0}$  with
$N(I_\up;I_\dn)=r$ and
${g}(I_\up^\prime;I_\dn^\prime)$ 
with $N(I_\up^\prime;I_\dn^\prime)=r+1$.   
This motivates us to decompose $F[\Phi]$ as
\begin{equation}
\label{eq:decomposition of F}
 F[\Phi] = F^\prime[\Phi]+\sum_{r=0}^{2\nu-1} F_r[\Phi]\,,
\end{equation}
where
\begin{equation}
\label{eq:Fr}
 F_r[\Phi]=\sumtwo{I_\up,I_\dn\subset \calI\backslash\{0\}}
                  {N(I_\up;I_\dn)=r+1}
                  |g(I_\up;I_\dn)|^2 
          -\sumtwo{I_\up,I_\dn\subset \calI\backslash\{0\}}
                  {N(I_\up;I_\dn)=r}
                           |\g{0}{0}|^2 
                                         \,, 
\end{equation}
and 
\begin{equation}
 \label{eq:Fprime}
 F^\prime[\Phi] = \sumtwo{I_\up,I_\dn\subset \calI\backslash\{0\}}
                         {N(I_\up;I_\dn)=0}|g(I_\up;I_\dn)|^2\,.
\end{equation}
The term $F^\prime[\Phi]$ is
apparently non-negative,
and therefore $F[\Phi]\ge0$ is implied by $F_r[\Phi]\ge0$ 
for $r=0,\dots,2\nu-1$.

We will prove $F_r[\Phi]\ge0$ by using~\eqref{eq:final condition}.
First we note that, for a pair of $I_\up$ and $I_\dn$ such that
$N(I_\up;I_\dn)=r$, the number of non-zero $\tilde{g}$ 
in the right-hand side of~\eqref{eq:final condition} is at most 
$2\nu-r-1$ (see the definition \eqref{eq:tg1} of $\tilde{g}$).
For such a pair of $I_\up$ and $I_\dn$, using the Schwarz inequality, we have
\begin{eqnarray}
\label{eq:inequality for tilde g 1}
|\tg{0}{0}|^2  
         & \le & \frac{2\nu-r-1}{\nu^2}\sum_{l\in \calD}
                    |\tg{l}{l}|^2 \nonumber\\ 
            &&      +\frac{2\nu-r-1}{(\nu+1)^2}
                 \sumJo
                    |\tg{k}{-k}|^2.
\end{eqnarray}
Then,
noting that 
$0 \le \frac{2\nu-r-1}{(\nu+1)^2}\le\frac{2\nu-r-1}{\nu^2}$, 
we find that
\begin{eqnarray}
\label{eq:Fr lower bound}
\sumtwo{I_\up,I_\dn\subset \calI\backslash\{0\}}
       {N(I_\up;I_\dn)=r}
       |\g{0}{0}|^2
& = &  
\sumtwo{I_\up,I_\dn\subset \calI\backslash\{0\}}
       {N(I_\up;I_\dn)=r}
       |\tg{0}{0}|^2 \nonumber\\
& \le & \frac{2\nu-r-1}{\nu^2}
       \sumtwo{I_\up,I_\dn\subset \calI\backslash\{0\}}
              {N(I_\up;I_\dn)=r}
              \sum_{l\in \calD}
                    |\tg{l}{l}|^2 \nonumber\\ 
&&    +\frac{2\nu-r-1}{(\nu+1)^2}
        \sumtwo{I_\up,I_\dn\subset \calI\backslash\{0\}}
               {N(I_\up;I_\dn)=r}
                 \sumJo
                    |\tg{k}{-k}|^2 \nonumber \\
& \le & \frac{(2\nu-r-1)(r+1)}{\nu^2}
       \sumtwo{I_\up,I_\dn\subset \calI\backslash\{0\}}
       {N(I_\up;I_\dn)=r+1}
       |g(I_\up;I_\dn)|^2\,.
\end{eqnarray}
To get the final line, we have used the fact that,
for a pair of $I_\up$ and $I_\dn$ with $N(I_\up;I_\dn)=r+1$,
there are $(r+1)$ elements $n$ in $\calD\cup(\calJ\backslash\{0\})$
for which we can find a suitable pair $I^\prime_\up$ and $I^\prime_\dn$ with
$N(I^\prime_\up;I^\prime_\dn)=r$ such that either
$\{n\}\cup I^\prime_\up = I_\up$ and $\{n\}\cup I^\prime_\dn = I_\dn$
or
$\{n\}\cup I^\prime_\up = I_\up$ and $\{-n\}\cup I^\prime_\dn = I_\dn$
holds.
Therefore we obtain
\begin{equation}
\label{eq:Fr ge 0}
F_r[\Phi] \ge \left( 1-\frac{(2\nu-r-1)(r+1)}{\nu^2} \right) 
               \sumtwo{I_\up,I_\dn\subset \calI\backslash\{0\}}
                      {N(I_\up;I_\dn)=r+1}
               |{g}(I_\up;I_\dn)|^2\,.
\end{equation}
Since $\frac{(2\nu-r-1)(r+1)}{\nu^2}\le1$,
we conclude that $F_r[\Phi]\ge0$.

Finally, we examine the condition for equality $F[\Phi] = 0$%
%
\footnote{%
We could easily find the condition for this equality
in the case of the kagome lattice~\cite{TanakaUeda2003}, 
but we have to consider this carefully in the present model.
The condition for the equality depends on 
the form of the finite energy condition~\eqref{eq:finite energy condition}.
This point should be kept in mind in applying the present method 
to other models. 
}.
Since $F^\prime[\Phi]\ge0$ by the definition 
and we have shown $F_r[\Phi]\ge 0$,
$F[\Phi]=0$ holds only when $F^\prime[\Phi]$ and $F_r[\Phi]$ are vanishing. 
Here it is easy to see that $F^\prime[\Phi]=0$ holds only if all the coefficients  
appearing in the sum in the right-hand side of~\eqref{eq:Fprime}, 
i.e., coefficients $g(I_\up;I_\dn)$ for $I_\up$ and $I_\dn$ such that 
$0 \notin I_\up\cup I_\dn$ and $N(I_\up;I_\dn)=0$,
are identically zero. 

To see when $F_r[\Phi]=0$ holds, 
let us look into the inequality~\eqref{eq:Fr ge 0}.
The factor $\frac{(2\nu-r-1)(r+1)}{\nu^2}$ in~\eqref{eq:Fr ge 0} is equal to 1
if $r=\nu-1$  and is less than 1 otherwise.
Therefore we find that $F_r[\Phi]$ with $r\ne \nu-1$
is vanishing only if all the coefficients $g(I_\up;I_\dn)$ 
in the sum in the right-hand side 
of~\eqref{eq:Fr} are zero.
On the other hand,
we need further consideration for $r=\nu-1$.
By using the inequality ~\eqref{eq:inequality for tilde g 1} with $r=\nu-1$,
we obtain
\begin{eqnarray}
\label{eq:r=nu-1}
\sumtwo{I_\up,I_\dn\subset \calI\backslash\{0\}}
       {N(I_\up;I_\dn)=\nu-1}
       |\g{0}{0}|^2
& \le & \frac{1}{\nu}
       \sumtwo{I_\up,I_\dn\subset \calI\backslash\{0\}}
              {N(I_\up;I_\dn)=\nu-1}
              \sum_{l\in \calD}
                    |\tg{l}{l}|^2 \nonumber\\ 
&&    +\frac{\nu}{(\nu+1)^2}
        \sumtwo{I_\up,I_\dn\subset \calI\backslash\{0\}}
               {N(I_\up;I_\dn)=\nu-1}
                 \sumJo
                    |\tg{k}{-k}|^2.
\nonumber\\
\end{eqnarray}
The right-hand side is rewritten as%
\footnote{
Note that $|I_\up\cap I_\dn\cap\calD|=n$ implies $|I_\up\cap {\bar{I}_\dn}\cap(\calJ\backslash\{0\})|=\nu-n$
when $N(I_\up;I_\dn)=\nu$.
Then, considering in a similar way below~\eqref{eq:Fr lower bound}, we obtain $R_n$.
}
\begin{equation}
 \sum_{n=0}^{\nu}
  \sumtwo{I_\up,I_\dn\subset \calI\backslash\{0\}}
  {N(I_\up;I_\dn)=\nu,~|I_\up\cap I_\dn\cap\calD|=n} 
  R_n |{g}(I_\up;I_\dn)|^2 
\end{equation} 
with $R_n=n/\nu+(\nu-n)\nu/(\nu+1)^2$.
Noting that $R_\nu=1$ we obtain
\begin{equation}
 F_{\nu-1}[\Phi]\ge\sum_{n=0}^{\nu-1} \left(1-R_n\right)  
  \sumtwo{I_\up,I_\dn\subset \calI\backslash\{0\}}
  {N(I_\up;I_\dn)=\nu,~|I_\up\cap I_\dn\cap\calD|=n} 
  |{g}(I_\up;I_\dn)|^2. 
\end{equation}
Now suppose that $F_{\nu-1}[\Phi]=0$.
Since $R_n<1$ for $n=0,\dots,\nu-1$,
we find that 
all the coefficients ${g}(I_\up;I_\dn)$ in the sum in the right-hand side
of the above inequality should be zero. 
We thus have $\tilde{g}(k,I_\up;-k,I_\dn)=0$ 
with $k\in\calJ\backslash\{0\}$ 
for any $I_\up,I_\dn\subset \calI\backslash\{0\}$ such that $N(I_\up;I_\dn)=\nu-1$%
\footnote{%
\label{fn:memo}
Put $I_\up^\prime=\{k\}\cup I_\up$ and $I_\dn^\prime=\{-k\}\cup I_\dn$. 
If $k\in I_\up$ or $-k\in I_\dn$, we have $\tilde{g}(k,I_\up;-k,I_\dn)=0$ by the definition; 
otherwise, since $N(I_\up^\prime;I_\dn^\prime)=\nu$ and 
$|I_\up^\prime \cap I_\dn^\prime \cap \calD|\ne\nu$, we have $g(I_\up^\prime;I_\dn^\prime)=0$, which
implies $\tilde{g}(k,I_\up;-k,I_\dn)=0$.
}.
Then, \eqref{eq:final condition} becomes
\begin{equation}
\label{eq:new condition on tg}
\tg{0}{0}
 =  -\frac{1}{\nu}
              \sum_{l\in \calD}
                    \tg{l}{l}
\end{equation} 
for $N(I_\up;I_\dn)=\nu-1$. Note that the number of non-zero $\tilde{g}$ 
in the right-hand side of \eqref{eq:new condition on tg} is at most one%
\footnote{%
Put $I_\up^\prime=\{l\}\cup I_\up$ and $I_\dn^\prime=\{l\}\cup I_\dn$. 
If $I_\up^\prime \cap I_\dn^\prime \cap \calD\ne\calD$, 
we have $\tilde{g}(l,I_\up;l,I_\dn)=0$ (see also footnote \ref{fn:memo}).
If $I_\up^\prime \cap I_\dn^\prime \cap \calD=\calD$, $|I_\up \cap I_\dn \cap \calD|=\nu-1$ 
and there is only one element $l$ in $\calD$ for 
which $\tg{l}{l}$ can be non-zero.   
}.
Repeating the same argument as above with the relation \eqref{eq:new condition on tg}, 
we find that 
$F_{\nu-1}[\Phi]=0$ 
only when all the coefficients in the sum in the right-hand side 
of \eqref{eq:Fr} 
are zero.   
Therefore, we conclude that $F[\Phi]=0$ holds 
only if $g(I_\up;I_\dn)=0$ for any pair of 
$I_\up$ and $I_\dn$ such that $0\in I_\up\cap I_\dn$ 
and
$0\notin I_\up\cup I_\dn$.  

So far we have proved that $E[\Phi]\ge -2\nu s$ and that the minimum
expectation value $-2\nu s$ is attained by states $\Phi$ 
which satisfy finite energy state 
condition~\eqref{eq:finite energy condition} and can be expanded as
\begin{equation}
 \Phi = \sumtwo{I_\up,I_\dn\subset \calI}
               {0\in I_\up\cup I_\dn,0\notin I_\up\cap I_\dn}
        g(I_\up;I_\dn)\Phi(I_\up;I_\dn).
\end{equation}
It follows {from} the above expression that 
any state which attains
the minimum expectation value $-2\nu s$ is
an eigenstate of $h_0$ as well as 
${-s\sum_{\sigma}a_{0,\sigma}^\dagger a_{0,\sigma}}$.
It is noted that
there really exist such eigenstates 
under condition~\eqref{eq:finite energy condition};
$d_{0,\up}^\dagger{\Phi}_0$ and 
$\left(\prod_{i\in \calI}d_{i,\up}^\dagger\right){\Phi}_0$
are examples.

By the continuity of eigenvalues 
we conclude that $-2\nu s$  is 
the minimum eigenvalue of $h_0$ for sufficiently large 
values of $t/s$ and
$U/s$, and that
the corresponding eigenstates are the same as those 
which give the minimum expectation value of $h_0$ 
in the limit $t/s,U/s\to\infty$.
It is easy to check that such eigenstates have
the properties stated in Lemma~\ref{lemma:main}.
Because of the translation invariance,
the same holds for all $\halph$ and
this completes 
the proof of Lemma~\ref{lemma:main}.        
\section{Proof of {Theorem~\ref{th:main}} for $s>0$; nearly-flat-band ferromagnetism}
\label{s:proof of theorem}
Suppose that the number $\Ne$ of electrons is $|\calV|$ and 
that the values of $t/s$ and $U/s$
are so large that Lemma~\ref{lemma:main} holds.
We note that Lemma~\ref{lemma:main} is associated only 
with the local Hamiltonians 
and that how large $t/s$ and $U/s$
should be is independent of 
the size of the whole lattice.

Since the minimum eigenvalue of 
the local Hamiltonians $\halph$ 
is $-2\nu s$, as claimed in Lemma~\ref{lemma:main}, 
the eigenvalue of 
the Hamiltonian $H=\sum_{\alpha\in \calV}\halph$
is bounded {from} below by $-2\nu s|\calV|$.
On the other hand, taking the fully saturated ferromagnetic state
\begin{equation}
 \label{eq:ferro state}
 \Phi_\mathrm{ferro} 
 = \left(\prod_{\alpha\in \calV}a_{\alpha,\up}^\dagger\right)\Phi_0
\end{equation}
as a variational state for $H$ and noting that
\begin{equation}
 -s a_{\beta,\dn}^\dagger a_{\beta,\dn}\Phi_\mathrm{ferro} = 0 
\end{equation}
and
\begin{equation}
 -s a_{\beta,\up}^\dagger a_{\beta,\up}\Phi_\mathrm{ferro} 
 = -s(2\nu-a_{\beta,\up} a_{\beta,\up}^\dagger)\Phi_\mathrm{ferro}
 =-2\nu s \Phi_\mathrm{ferro}  
\end{equation}
for all $\beta\in \calV$,
we find that
the upper bound on the ground state energy
is also given by $-2\nu s|\calV|$.
Therefore, 
the ground state energy 
is exactly $-2\nu s|\calV|$. 
It is easy to see that the state $\Phi_\mathrm{ferro}$ has
the maximal total spin $\Stot=\Ne/2$.
Our remaining task is to prove 
the uniqueness of the ground state up to $(\Ne+1)$-fold degeneracy
due to the spin rotation symmetry. 

Let $\PhiG$ be a ground state,
and assume that
$\PhiG$ is expanded in terms of the basis states 
$\Phi(V_\up,B_\up;V_\dn,B_\dn)$ in~\eqref{eq:basis}
with $\Ne=|\calV|$. 
Since $H\PhiG=-2\nu s|\calV|\PhiG$, $\PhiG$ must satisfy 
$\halph\PhiG = -2\nu s \PhiG$ for all $\alpha\in \calV$.
Thus $\PhiG$ must satisfy the properties stated in Lemma~\ref{lemma:main}.

The condition~\eqref{eq:condition1 in Lemma} implies
that
\begin{equation}
 a_{\alpha,\up}^\dagger a_{\alpha,\dn}^\dagger\PhiG=0
\end{equation} 
for any $\alpha\in\calV$.
Since $a_{\alpha,\up}^\dagger a_{\alpha,\dn}^\dagger\Phi(V_\up,B_\up;V_\dn,B_\dn)\ne0$
if $\alpha\notin V_\up\cup V_\dn$ and the states of the form 
$a_{\alpha,\up}^\dagger a_{\alpha,\dn}^\dagger\Phi(V_\up,B_\up;V_\dn,B_\dn)$ are 
linearly independent with each other,
$\Phi$ is expanded only in terms of the basis states $\Phi(V_\up,B_\up;V_\dn,B_\dn)$
with $V_\up \cup V_\dn=\calV$. 
Then, taking into account $\Ne=|\calV|$, 
we find that $\PhiG$ is written as
\begin{equation}
\label{eq:PhiG spin representation}
 \PhiG = \sum_{\{\sigma\}} \varphi(\{\sigma\})
         \left(\prod_{\alpha\in \calV}
           a_{\alpha,\sigma_\alpha}^\dagger
	   \right)\Phi_0 \,,
\end{equation}
where $\{\sigma\}$ is a shorthand for a spin configuration
$\{\sigma_\alpha\}_{\alpha\in \calV}$ of electrons each of 
which is singly occupying a state corresponding to 
${a}_{\alpha,\sigma_\alpha}$, the summation 
is taken over all spin configurations, and $\varphi(\{\sigma\})$
is a new complex coefficient. 
As for the product of fermion operators, we adopt the same rule as
in~\eqref{eq:d basis} with another one-to-one mapping $\theta^\prime$ from $\calV$ to $\Bbb{Z}$.

Let us impose 
the condition~\eqref{eq:condition2 in Lemma} on $\PhiG$ 
in the form of~\eqref{eq:PhiG spin representation}.
Note that \eqref{eq:condition2 in Lemma} must hold for sites in all $\La_\alpha$,
i.e., for all sites in $\La$. 
Choose site $x=m(\beta,\gamma)$.
By the definition of $a_{\alpha,\sigma}$, one finds that
$\{c_{m(\beta,\gamma)},a_{\alpha,\sigma}^\dagger\}$ is 1 if $\alpha$ is either $\beta$ or $\gamma$,
and zero otherwise.
Then, we have 
\begin{eqnarray}
\lefteqn{c_{m(\beta,\gamma),\dn}c_{m(\beta,\gamma),\up}\PhiG} \nonumber\\
&=&
  \sum_{\{\sigma\}}
  \mathsf{S}_{\calV}^\beta \mathsf{S}_{\calV \backslash\{\beta\} }^\gamma
  \varphi(\{\sigma\})
  c_{m(\beta,\gamma),\dn}c_{m(\beta,\gamma),\up}
  a_{\beta,\sigma_\beta}^\dagger a_{\gamma,\sigma_\gamma}^\dagger
  \left(
   \prod_{\alpha\in \calV\backslash\{\beta,\gamma\}}
              {a}_{\alpha,\sigma_\alpha}^\dagger
  \right)\Phi_0 
  \nonumber\\
&=&
  \sum_{\{\sigma\}}
  \mathsf{S}_{\calV}^\beta \mathsf{S}_{\calV \backslash\{\beta\} }^\gamma
  \varphi(\{\sigma\})
  \left\{  
  \chi[\sigma_\beta=\up,\sigma_\gamma=\dn]
  -
  \chi[\sigma_\beta=\dn,\sigma_\gamma=\up]
  \right\}  
  \left(
   \prod_{\alpha\in \calV\backslash\{\beta,\gamma\}}
              {a}_{\alpha,\sigma_\alpha}^\dagger
  \right)\Phi_0 
  \nonumber\\
\end{eqnarray}   
where $\mathsf{S}_{\calV}^\beta$ and $\mathsf{S}_{\calV \backslash\{\beta\}}^\gamma$ 
are defined similarly as in~\eqref{eq:sign factor S} 
with $\theta$ replaced by $\theta^\prime$.
Since all the states 
$\left(
   \prod_{\alpha\in \calV\backslash\{\beta,\gamma\}}
              {a}_{\alpha,\sigma_\alpha}^\dagger
  \right)\Phi_0$ 
in the last line are linearly independent,
it follows from 
the condition $c_{m(\beta,\gamma),\dn}c_{m(\beta,\gamma),\up}\PhiG=0$ 
that
$
  \varphi(\{\sigma\})=\varphi(\{\tau\})
$
for any  pair of spin configurations
$\{\sigma\}$ and $\{\tau\}$ such that
$\sigma_\beta=\tau_\gamma, \sigma_\gamma=\tau_\beta$, 
and $\sigma_\alpha=\tau_\alpha$ for $\alpha\ne\beta,\gamma$.
 
Repeating the same argument for all sites in $\La$,
we find that
\begin{equation}
\label{eq:condition on coefficients}
 \varphi(\{\sigma\}) = \varphi(\{\tau\})
\end{equation} 
if  
$\sum_{\alpha\in \calV}\sigma_\alpha=\sum_{\alpha\in \calV}\tau_\alpha$
(we regard $\up$ as +1 and $\dn$ as $-1$ in the sum).  
Therefore $\PhiG$ is always expanded as
\begin{equation}
 \PhiG = \sum_{M=-\Ne}^{\Ne}\varphi_M (S_\mathrm{tot}^-)^{\Ne-M}
         \Phi_\mathrm{ferro}
\end{equation}
with complex coefficients $\varphi_M$. 
This completes the proof of Theorem~\ref{th:main} for $s>0$.
\jour{
\bigskip\\
{\small
\textbf{Acknowledgements}~~I would like to thank Hal Tasaki for valuable discussions.The present work was supported by JSPS Grants-in-Aid for Scientific Research no.~25400407.}
}
{
\begin{acknowledgements}
I would like to thank Hal Tasaki for valuable discussions.The present work was supported by JSPS Grants-in-Aid for Scientific Research no.~25400407.
\end{acknowledgements}
}
\begin{appendices}
\renewcommand{\theequation}{\thesection.\arabic{equation}}
\setcounter{equation}{0}
\section{Proof of {Theorem~\ref{th:main}} for $s=0$; flat-band ferromagnetism}
\label{s:flat-band ferro}
Throughout this section, we  assume that $s=0$
and the other parameters $t$ and $U$ take
arbitrary positive values.

As we have seen in section~\ref{s:single electron},
the single-electron ground state energy is zero
and $|\calV|$-fold degenerate for $s=0$, and the single-electron
ground states 
are given by $\aasd\Phi_0$ with $\alpha\in \calV$. 
Since both $\Hhop$ with $s=0$ and $\Hint$ are positive semidefinite,
we find that the energy eigenvalue of $H$ is bounded from below by zero.
Taking a state $\Phi_\mathrm{ferro}$ in \eqref{eq:ferro state}
as a variational state, we conclude that the ground state energy of $H$ for $\Ne=|\calV|$
is exactly zero.

To show the uniqueness of the ground state, 
we will use the following property which was pointed out by Mielke~\cite{Mielke93}. 
Consider Hubbard models with $M$-fold degenerate 
single-electron ground state energy.
Then, we can construct 
a set of $M$ single-electron ground states 
which satisfy the following condition.
Let $\Phi_{1,\sigma}^{(i)}$ with
$i=1,2,\dots,M$ be linearly independent 
single-electron ground states with spin $\sigma$. 
For each $i$, there exists site $x_i$ such
that $c_{x_i,\sigma}\Phi_{1,\sigma}^{(i)} \ne 0 $ and
$c_{x_i,\sigma}\Phi_{1,\sigma}^{(j)}=0$ for $j\ne i$. 

In our model, we can construct single-electron ground states
which possess the above property in the following manner. 
For each $\alpha\in {\calV}$, let us 
define a new fermion operator by
\begin{equation}
 \bar{a}_{\alpha,\sigma} 
= a_{\alpha,\sigma}
  -\sum_{n=1}^{L-1}(-1)^{n}a_{\alpha+n\delta_\nu,\sigma}
\end{equation}
(recall that $\calV$ is periodic). Since $\bar{a}_{\alpha,\sigma}$ is a linear combination
of $a_{\beta,\sigma}$ with $\beta\in \calV$,
the single-electron states $\bar{a}_{\alpha,\sigma}^\dagger\Phi_0$
apparently have energy 0.
Furthermore, we have 
\begin{equation}
\label{eq:c bar a}
 \acom{\bar{a}_{\beta,\sigma}^\dagger,c_{m(\alpha,\alpha+\delta_\nu),\sigma}}= 
 \left\{
  \begin{array}{@{\,}ll}
   2 & \mbox{if $\beta=\alpha$;}\\
   0   & \mbox{otherwise.}
  \end{array}
 \right.
\end{equation}
These anticommutation relations imply that
the single-electron states $\bar{a}_{\alpha,\sigma}^\dagger\Phi_0$ 
with $\alpha\in \calV$ have the desired property.
We note that \eqref{eq:c bar a} also implies 
the linear independence of these single-electron states. 

Let us prove the uniqueness of the ground state
by using $\{\bar{a}_{\alpha,\sigma}^\dagger\}_{\alpha\in{\calV}}$
introduced as above.
Let $\PhiG$ be a ground state, which must be 
a zero energy state for both $\Hhop$ and $\Hint$.
Since $\PhiG$ is a zero energy state for $\Hhop$, we expand it as
\begin{equation}
 \PhiG = \sumtwo{V_\up,V_\dn\subset \calV}
                {|V_\up|+|V_\dn|=|\calV|}
		\varphi^\prime(V_\up;V_\dn)
		\left(
		 \prod_{\alpha\in V_\up}
		\bar{a}_{\alpha,\up}^\dagger
		\right)
		\left(
		\prod_{\alpha\in V_\dn}
		\bar{a}_{\alpha,\dn}^\dagger
		\right)
		\Phi_0
\label{eq:flat-band PhiG}
\end{equation}
with complex coefficients $\varphi^\prime(V_\up;V_\dn)$.
To be a zero energy state of the on-site interaction, $\PhiG$  
in the form of~\eqref{eq:flat-band PhiG} 
must further satisfy the condition $c_{x,\dn}c_{x,\up}\PhiG=0$ for all $x$ in $\La$.
We examine this condition for sites $x=m(\alpha,\alpha+\delta_\nu)$ with $\alpha\in\calV$.
Then, taking account of \eqref{eq:c bar a},
we find that $\varphi^\prime(V_\up;V_\dn)$ is vanishing if $V_\up\cap V_\dn\ne\emptyset$. 
A ground state $\PhiG$ is thus rewritten as
\begin{equation}
 \PhiG = \sum_{\{\sigma\}} \varphi^\prime(\{\sigma\}) 
  \left(
	  \prod_{\alpha \in \calV} 
	  \bar{a}_{\alpha,\sigma_\alpha}^\dagger
		 \right)
	 \Phi_0
\end{equation}
with new coefficients $\varphi^\prime(\{\sigma\})$. 
Now we have expressed $\PhiG$ 
in the same fashion as in~\eqref{eq:PhiG spin representation}.
Examining repeatedly the condition $c_{x,\dn}c_{x,\up}\PhiG=0$ for sites $x=m(\beta,\beta+\delta_l)$ 
with $l\ne\nu$,
we obtain the same relation for $\varphi^\prime(\{\sigma\})$ 
as that in~\eqref{eq:condition on coefficients}. 
Therefore, we conclude that the ground state is unique apart from the degeneracy due to 
the spin rotation symmetry.
This completes the proof of Theorem~\ref{th:main} for $s=0$.
\end{appendices}

\end{document}